# Equilibrium Clusters in Concentrated Lysozyme Protein Solutions

P. Kowalczyk[*1], A. Ciach[2], P. A. Gauden[3], A. P. Terzyk[3]

*(1) Nanochemistry Research Institute, Department of Chemistry, Curtin University of Technology, P.O. Box U1987, Perth, 6845 Western Australia, Australia.*

*(2) Institute of Physical Chemistry, Polish Academy of Science, Kasprzaka Street 44/52, 01-224 Warsaw, Poland.*

*(3) Department of Chemistry, Physicochemistry of Carbon Materials Research Group, N. Copernicus University, Gagarin St. 7, 87-100 Torun, Poland.*

**Abstract:** We have studied the structure of salt-free lysozyme at 293 K and pH 7.8 using molecular simulations and experimental SAXS effective potentials between proteins at three volume fractions, $\phi = 0.012$, $0.033$, and $0.12$. We found that the structure of lysozyme near physiological conditions strongly depends on the volume fraction of proteins. The studied lysozyme solutions are dominated by monomers only for $\phi \leq 0.012$; for the strong dilution 70% of proteins are in a form of monomers. For $\phi = 0.033$ only 20% of proteins do not belong to a cluster. The clusters are mainly elongated. For $\phi = 0.12$ almost no individual particles exits, and branched, irregular clusters of large extent appear. Our simulation study provides new insight into the formation of equilibrium clusters in charged protein solutions near physiological conditions.

# 1. Introduction

Spatial distribution of proteins in physiological solutions is very important for living matter; individual particles or clusters of particles can be associated with quite different properties or functions. In particular, a change of structural properties may lead to different diseases [1-14]. For this reason it is important to understand how different factors influence the structure of proteins in different solutions. In 2004, Stradner et al. [15] published the fundamental work connected with self-assembly into equilibrium clusters in systems containing weakly charged globular proteins as well as in systems containing weakly charged colloidal particles. In the case of colloids, confocal microscopy experiments showed directly the presence of equilibrium clusters of different sizes. Campbell et al. [16] also observed cluster formation in system composed of weakly charged colloids and non-adsorbing polymers, whose radius of gyration was much smaller than the radius of colloidal particles. Effective interaction between the particles in this system consists of hard-core repulsion at distances shorter than the particle diameter, strong short-range attraction and weak long-range repulsion (SALR) [17-20]. The repulsion originates from screened electrostatic interactions, and the attraction results from depletion interactions induced by the polymer. Careful analysis of three-dimensional fluorescence confocal microscope images showed that colloids form stable equilibrium clusters at low volume fractions [16]. Upon increasing the volume fraction of colloids the clusters grow in size and become increasingly anisotropic, until finally a network of clusters is formed (e.g. stable gel structure consisting of Bernal spirals [16,21,22]).

In concentrated salt-free solutions of lysozyme close to physiological pH at 278, 288, and 298 K, cluster formation was concluded on the basis of small angle X-ray (SAXS) and neutron scattering (SANS) experiments by Stradner et al. [15,23]. The authors predicted an increase of average cluster size with increasing volume fraction of lysozyme in analogy with colloidal systems [15,16,23]. Neutron spin echo (NSE) and SANS experiments due to Porcar et al. [24] clearly showed that at high-volume fractions lysozyme proteins self-assemble into so-called dynamic clusters which have a finite lifetime, and their number and shape fluctuate dramatically. Thus, the equilibrium protein clusters resemble the structure of living polymers [25-28]. In contrast, at low-volume fractions of lysozyme the system is dominated by individual proteins [15,23,24]. Porcar et al. [24] argued that the observed clusters are ergodic and the macroscopic properties of the long time limit are still determined by monomeric proteins. In simulation and theoretical studies the formation of equilibrium clusters and gelation were found for high enough concentrations for several model systems characterized by different versions of the SALR potential [29-34]. Moreover, ordering of the clusters into periodic structures, similar to those observed in amphiphilic systems, was theoretically predicted for low temperatures [20,33]. The authors of the above mentioned papers argued that the competition between short-range attraction and long-range repulsion between weakly charged particles is a sole of equilibrium cluster formation in concentrated protein as well as in colloidal solutions [20,29-34].

The experimental and theoretical results quoted above seem to suggest that there is consensus concerning the analogy between cluster formation in systems composed of globular weakly charged proteins, and colloids in salt-free solvents. However, the results of Stradner et al. [15] have been recently questioned by the outcome of a joint SANS/SAXS study performed at very similar thermodynamic conditions. Shukla et al. [35] definitely excluded the existence of an equilibrium cluster phase and claimed that in the concentrated salt-free lysozyme solutions the effective repulsive forces keep the individual lysozyme proteins apart. The main argument of Shukla et al. [35] was that the position of low-angle interference peak in both SANS and SAXS scattering patterns depends on the concentration of lysozyme at the studied physiological conditions. This observation contradicts the earlier investigations, because according to Stradner et al. [15] the cluster-cluster peak position is concentration-independent. Later Stradner et al. [36] admitted that the peak position indeed depends on density.

The experiments reported by Shukla et al.[35] were performed independently by several groups and using different methods, therefore should be taken into account seriously. Thus, *one can conclude that the equilibrium cluster formation in salt-free lysozyme near physiological conditions is an open question*. Since the proteins are too small for direct observation, we believe that in order to reach more definite conclusions, extensive computer simulations for the interaction potential derived by Shukla et al. [35] from the results of scattering experiments should be performed. Shukla et al. [35] fitted the scattering patters by effective double Yukawa potentials with strong concentration/temperature dependence of short-range attraction and long-range repulsion amplitudes and ranges. If this potential is correct approximation, computer simulations should give similar insight as microscopic images. In the work reported in this letter we used molecular simulation techniques to investigate the properties of salt-free lysozyme at 293 K and pH 7.8.

**2. Computation Details**

Parallel tempering Monte Carlo method and cluster moves [37,38] were used; in all studied systems we simulated at least N = 3000-4200 proteins (e.g. box size of L = 91-146 nm). Parallel tempering involves conducting several Monte Carlo simulations of the same system at different temperatures simultaneously [38]. In the current work we simulated 10-20 replicas in N,V,T ensemble with temperatures exponentially distributed between 298 K and 500 K. The parameters of exponential distribution of temperatures and frequency of replica exchange were adjusted to $\approx 30$ % acceptance rate. In the N,V,T ensemble, the probability of replica exchange between two randomly selected replicas was expressed by Metropolis-Hastings rule [38],

$$P = \min\{1, \exp[(\beta_i - \beta_j)(U_i - U_j)]\} \tag{1}$$

where $U_i$ and $U_j$ are the potential energies of the current configurations of the simulations conducted at $\beta_i = (k_B T_i)^{-1}$ and $\beta_j = (k_B T_j)^{-1}$, $k_B$ denotes Boltzmann constant.

In each replica (i.e. independent N,V,T ensemble) the potential energy was equilibrated by combination of single particle displacements and cluster moves. The single displacement step was accepted according to standard Metropolis sampling scheme [37,38],

$$P = \min\{1, \exp[\beta_i \Delta U_i]\} \tag{2}$$

where $\Delta U_i$ denotes the change of the total potential energy of the $i-th$ replica randomly selected for perturbation. Displacement step size was adjusted to $\approx 40$ % acceptance rate. It is commonly known that single particle displacement is a very inefficient perturbation once the system of studied particles self-assembles into clusters. This is because the clusters are energetically stable structures. In real experiment, the energy barrier that separates individual monomers and clusters can be overcome by spontaneous thermal fluctuations. In computer experiment, such thermal fluctuations are rare events. To enhance the efficiency of sampling, in each replica we introduced simple cluster moves. First, the random particle (e.g. cluster seed), $\underline{r}_s = \{x_s, y_s, z_s\}$, is swapped. Next, we identify the particle, $\underline{r}_i = \{x_i, y_i, z_i\}$, which fulfills the following condition,

$$\|\underline{r}_s - \underline{r}_i\| < \xi \tag{3}$$

where $\|\underline{r}_s - \underline{r}_i\| = \sqrt{(x_s - x_i)^2 + (y_s - y_i)^2 + (z_s - z_i)^2}$ denotes Euclidean distance, and $\xi$ is the control parameter. In our calculations, $\xi = a \cdot \sigma$, $a \in [1, 1.2]$. Next, this process goes iteratively, considering all particles until the largest possible cluster is recorded, as is schematically presented in figure 1. Clearly, for same cases the cluster is consisting only of one monomer. Finally, the identified cluster is randomly displaced to a new position. The cluster displacement is accepted according to standard Metropolis probability [37,38]. As previously, the displacement step size was adjusted to $\approx 40$ % acceptance rate.

Effective potentials between lysozyme particles at 293 K and different volume fractions were taken from SANS/SAXS scattering experiments due to Shukla et al. [35] and are shown in figure 2.

## 3. Results and Discussion

Our results agree with experiments of Porcar et al. [24]. Let us first discuss the cluster size distribution computed from equilibrium configurations. We found that at the lowest volume fraction of $\phi = 0.012$ the protein solution is dominated by individual lysozyme particles. Around 70 % of lysozyme proteins are in a form of monomers, as is presented in figure 3 and in the movie attached to supplementary material. The remaining 30 % of proteins self-assemble into small equilibrium clusters, mainly dimers and trimers, but small amount of linear/compact tetramers and pentameters can be found too. In figure 3 the percent of particles belonging to a cluster composed of *N* particles, *p(N)*, is shown for $\phi = 0.012$ and $\phi = 0.033$. In figure 4 typical clusters taken from snapshots of equilibrium configurations are depicted for $\phi = 0.012$. An increase in volume fraction of lysozyme up to $\phi = 0.033$ changes the structure of concentrated protein solution significantly. The number of individual proteins drops to 20 % (see figure 5 and the movie attached to supplementary material). Moreover, cluster distribution function *p(N)* broadens, indicating polydispersity of equilibrium clusters, as is displayed in figure 3. Close inspection of equilibrium configurations indicates that lysozyme proteins are mainly self-assembled into elongated equilibrium clusters (e.g. living polymers as is shown in the movie attached to supplementary material), but shorter equilibrium clusters composed of tightly packed lysozyme (e.g. with topology similar to Bernal spiral [16,21,22]) appear too, as shown in figure 5. Further increase in protein concentration results in progressive growth of equilibrium clusters. According to our simulation results for $\phi = 0.12$, equilibrium lysozyme clusters are large and irregularly branched, as is presented in figure 6. Within those equilibrium clusters some proteins are tightly packed, but compact globular structures with radius much larger than the particle diameter are absent. It is very difficult to compute a reliable cluster distribution for $\phi = 0.12$, because the large equilibrium lysozyme clusters are influenced by the finite size of the simulation box; in the bulk the clusters can be larger. Nevertheless, the structure of the lysozyme clusters at the highest studied volume fraction agrees qualitatively with recent simulations of a similar model SALR potential by Toledano at. [29], and with experimental reports for colloids by Campbell at al. [16].

In addition to cluster size distribution we calculated the radial distribution function $g(r)$ describing the ratio between the local density of particles at a distance $r \equiv r \cdot \sigma^{-1}$ from a chosen particle, and the average density. In figure 7 $g(r)$ is shown for the three volume fractions ( $g(r)$ is shifted vertically by 1 and by 2 for $\phi = 0.033$ and $\phi = 12$, respectively). We shall label the systems for $\phi = 0.012$, $\phi = 0.033$ and $\phi = 0.12$ as I, II, and III, respectively. As usual, the radial distribution function is characterized by a sharp peak at $r \approx 1$, where the distance is measured in units of the diameter of the particle, $\sigma$. The first peak corresponds to increased probability of finding a pair of particles at contact, and results from short-range attraction. The maximum at contact is followed by a broad minimum where $g(r) < 1$, indicating depleted density around each particle. The distance from the first particle corresponding to depleted density is $r \leq r_d$, with $r_d \approx 3$ for the two lower concentrations and $r_d \approx 2.5$ for $\phi = 0.12$. This result indicates that majority of proteins are far apart from each other, but some form small clusters - dimers, triangles or tetrahedral - of radius $\sim \sigma$. For $\phi = 0.033$ and $\phi = 0.12$, however, we observe sharp peaks for $r \sim 1.5$ and $r \sim 2$, corresponding to discrete distances associated with local geometries of larger clusters (compact or linear). In case I small maximum for $r \sim 2$ signals small amount of larger clusters. The secondary maximum corresponding to cluster-cluster correlations is not clearly seen. The finite size of the simulation box influences the results for the correlation function at large distances, and we cannot draw definite conclusions concerning packing of clusters – this is not our goal here, however. We should note that simulations of a model system for volume fractions similar to our highest concentration yield very similar shape of $g(r)$, with characteristic broad minimum and several discrete peaks for $1.5 < r < 3$ (see figure 6 in reference [29]). The maximum corresponding to cluster-cluster correlation was observed by Toledano et al. [29] for low temperatures. Due to finite size effects we did not attempt to calculate the structure factor, since the peak position at low $k$ is strongly influenced by the size of the box.

In order to understand our results, let us note that in equilibrium the free energy F=U-TS should be at minimum. The entropy assumes maximum for random distribution of particles. In order to estimate the internal energy for random distribution of particles, let us first discuss the pair potential. The effective potential between the studied lysozyme proteins strongly depends on the volume fraction. The long-range repulsion with maximum around 1.7 $k_B T$ is observed for the lowest studied volume fraction, $\phi = 0.012$. As the volume fraction of lysozyme increases, the repulsion maximum is gradually shifted towards shorter distances between the particles, and its height decreases (see figure 2).

The average separation between individual particles is given by $r^* = v^{1/3}$, where the volume per particle is $v = V/N = \dfrac{\pi}{6\phi}$ in $\sigma^3$ units. For the considered volume fractions the average separation between individual particles takes the values $r_I^* = 3.521$, $r_{II}^* = 2.513$, and $r_{III}^* = 1.634$. It is instructive to compare the average separations between individual proteins and the position of the repulsive barrier for each volume fraction (see figure 2). In the first case the average distance is almost beyond the range of repulsion, while in the case III the average distance corresponds to substantial positive energy of a pair of first neighbours. More precisely, $V^*(r_I^*) = 0.049$, $V^*(r_{II}^*) = 0.054$, and $V^*(r_{III}^*) = 0.177$. The average energy per volume in each case can be roughly estimated by $e^* \approx \frac{z}{2}\frac{6\phi}{\pi}V^*(r^*)$ where $z$ is the average number of nearest neighbours of a given particle. For fcc-like structure z = 12. We obtain $e_I^* \approx 0.001\frac{z}{2}$, $e_{II}^* \approx 0.003\frac{z}{2}$, and $e_{III}^* \approx 0.041\frac{z}{2}$. Note that if the proteins would be uniformly distributed, then the energy per volume would increase 3 times when the volume fraction increases from $\phi = 0.012$ to $\phi = 0.033$ and it would increase 36 times when the volume fraction increases from $\phi = 0.012$ to $\phi = 0.12$. On the other hand, when a fraction of particles $f$ form tight dimers, then the corresponding contribution to the energy per volume is $\sim -3f\frac{6\phi}{\pi}$, where we estimated the energy of a dimer by -3, which is close to the average energy in the attractive part of the potential (see figure 2). Each pair of particles belonging to a very small, tight cluster contributes $\sim -3\frac{6\phi}{\pi}$ to the system energy per volume (when the distance between the particles in the pair is within the attraction range). For low volume fractions it is possible to assemble the molecules into small clusters, separated by distances larger than the repulsion range. In this case cluster formation leads to a substantial decrease in energy, at the cost of decreased entropy, i.e. -TS increases. Since U decreases and -TS increases when clusters are formed, F=U-TS assumes a minimum for a compromise, i.e. when some fraction of particles belong to a cluster. This fraction increases with increasing volume fraction of the proteins, because in this way substantial increase in energy that would occur for monomers is avoided. For still larger volume fractions small clusters would be separated by distances corresponding to repulsive part of the interaction energy. As a result, positive contribution to the energy coming from interactions between particles belonging to different clusters would occur. On the other hand, when the size of the cluster is equal or larger than the range of the repulsion barrier, some pairs of proteins belonging to the same cluster would yield a positive contribution to the energy. The number of such pairs is the smallest in a linear, and the largest in a compact, globular cluster. It is thus energetically favourable to grow the cluster in one dimension, and keep the distance between elongated clusters larger than the repulsion range. Further increase of concentration leads to formation of large, irregular clusters, which do not contain compact, globular parts, but rather resemble a network-like shape. *Note that in such complex situation the average separation between the objects (clusters or monomers), and as a result the peak position of the structure factor depend on concentration.* Thus, our results do not contradict the results of experiments reported by Shukla et al. [35].

## 4. Conclusions

In summary, we have studied the structure of salt-free lysozyme at 293 K and pH 7.8 using molecular simulations and experimental SAXS effective potentials between proteins at three volume fractions, $\phi = 0.012$, 0.033, and 0.12. We found that the studied lysozyme solutions are dominated by monomers only for $\phi \leq 0.012$; for the strong dilution 70% of proteins are in a form of monomers. For $\phi = 0.033$ only 20% of proteins do not belong to a cluster. The clusters are mainly elongated. For $\phi = 0.12$ almost no individual particles exits, and branched, irregular clusters of large extent appear. These results, obtained for the interaction potential derived by Shukla et al. [35] from scattering experiments, contradict the conclusion drawn by these authors that individual particles rather than clusters exist at all studied volume fractions. On the other hand, since we observe equilibrium between monomers and polydisperse clusters with the fraction of monomers depending strongly on the volume fraction, the structure is more complicated than initially concluded by Stradner et al. [15].

**Acknowledgments**

P.K. acknowledges partial support by the Office of Research & Development, Curtin University of Technology, Grant CRF10084. P.G. and A.P.T. acknowledge the use of the computer cluster at Poznan Supercomputing and Networking Centre as well as the Information and Communication Technology Centre of the Nicolaus Copernicus University (Torun, Poland). The work of AC was partially supported by the Polish Ministry of Science and Higher Education, Grant No. NN 202 006034.

**Supplementary material:** Snapshots of lysozyme at 293 K, pH 7.8, and different volume fractions of proteins: 0.012, 0.033, and 0.12 (movie lys1.mpg and lys2.mpg).

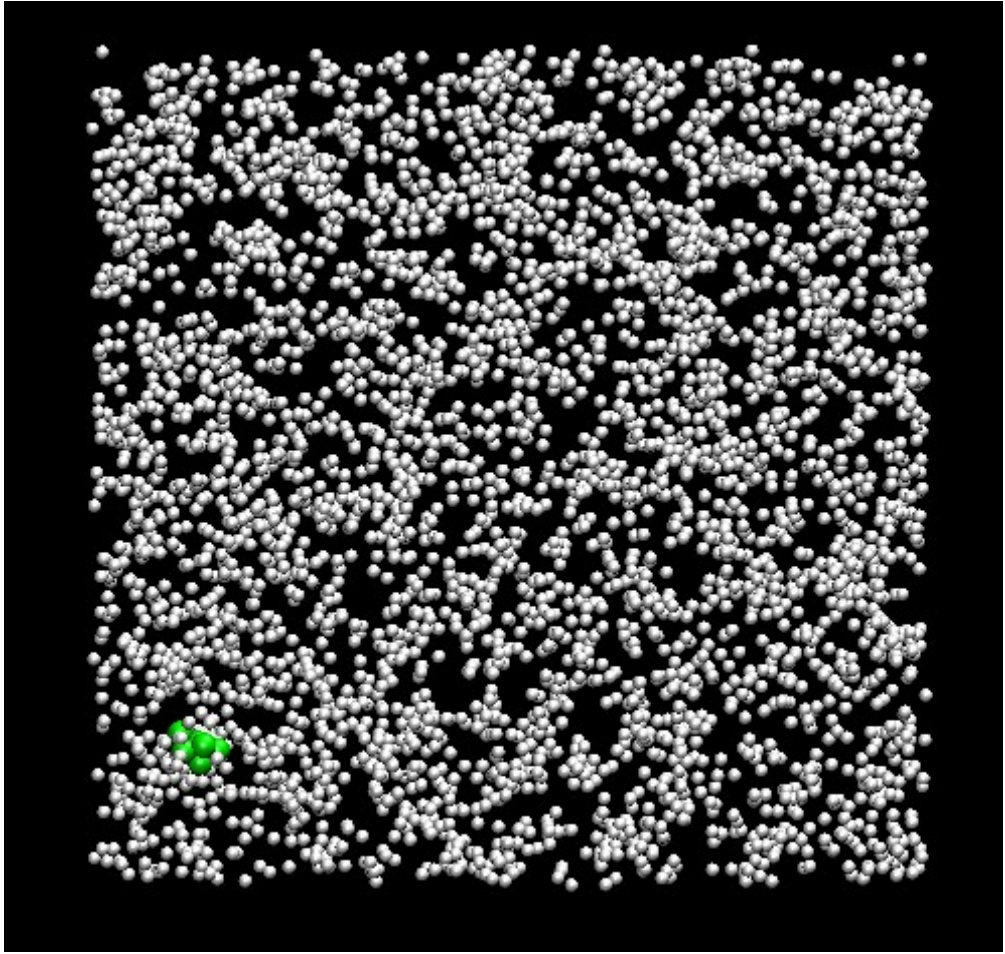

**figure 1.** Protein cluster (the hexamer marked by green spheres) automatically identified by the current algorithm.

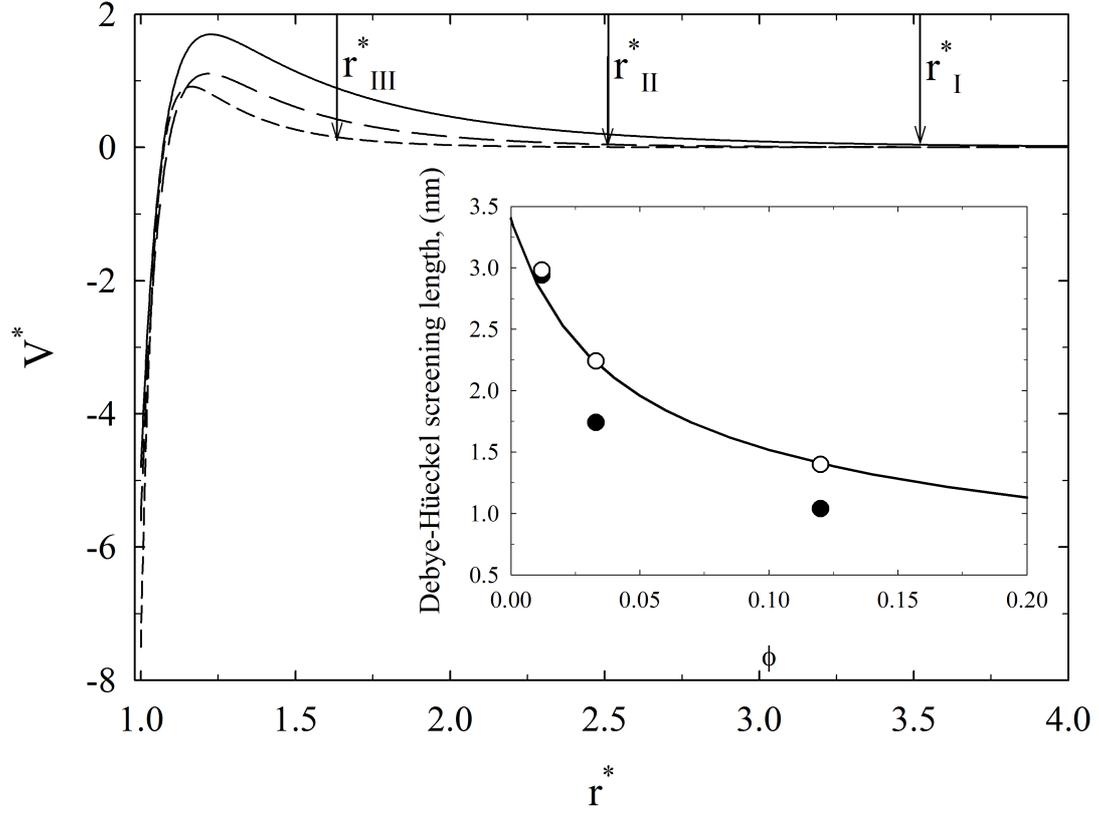

**figure 2.** The effective slat-free lysozyme-lysozyme potentials, $V^* = V \cdot \beta$, $r^* = r \cdot \sigma^{-1}, \beta = (k_B T)^{-1}$, $\sigma = 3.648$ nm, at 293 K, pH 7.8, and different volume fractions of proteins: 0.012 (solid lines), 0.033 (long-dashed lines), and 0.12 (short-dashed lines). The inner plot presents variation of Debye-Hüeckel screening length with volume fraction of lysozyme (solid line denotes present calculations and open circles are taken from Shukla et al. [35]). Note that the range of repulsive interactions (black circles) is comparable with Debye-Hüeckel screening lengths.

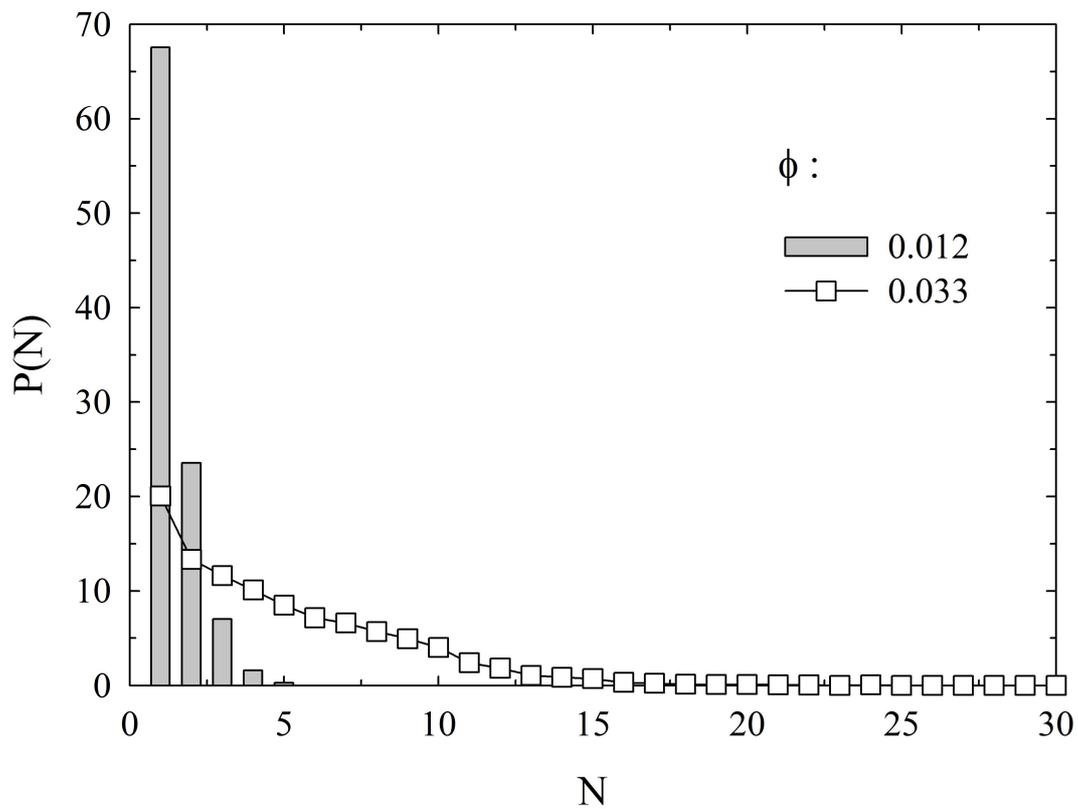

**figure 3.** Cluster size distributions computed for salt-free lysozyme at 293 K and pH 7.8. The volume fractions of proteins are displayed in the plot.

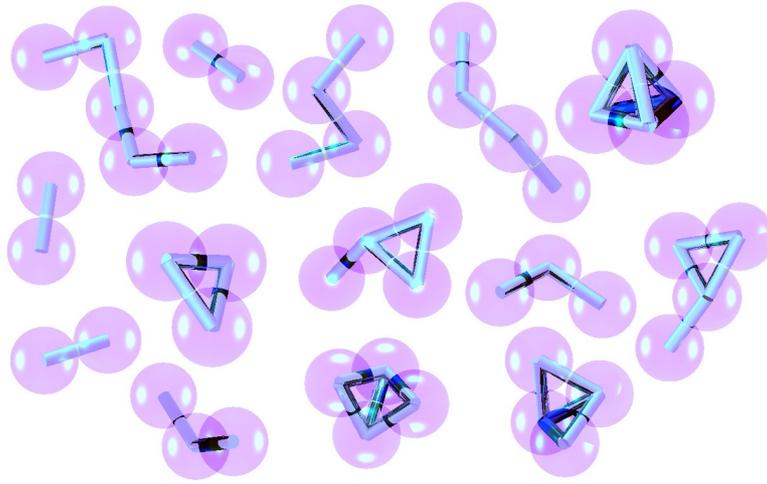

**figure 4.** Equilibrium clusters of lysozyme at 293 K and pH 7.8 and volume fraction of 0.012.

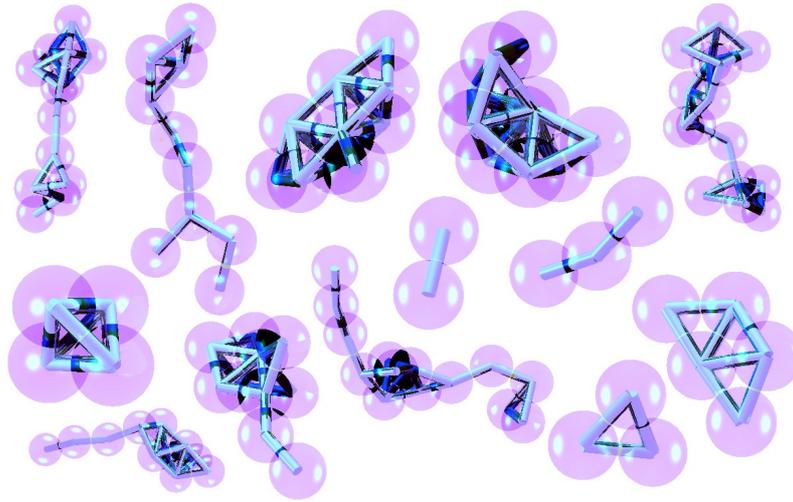

**figure 5.** Equilibrium clusters of lysozyme at 293 K and pH 7.8 and volume fraction of 0.033.

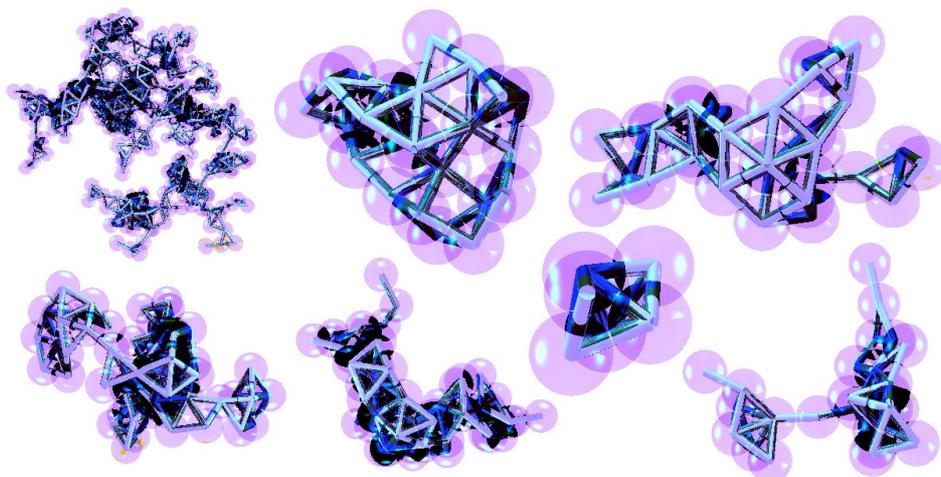

**figure 6.** Equilibrium clusters of lysozyme at 293 K and pH 7.8 and volume fraction of 0.12.

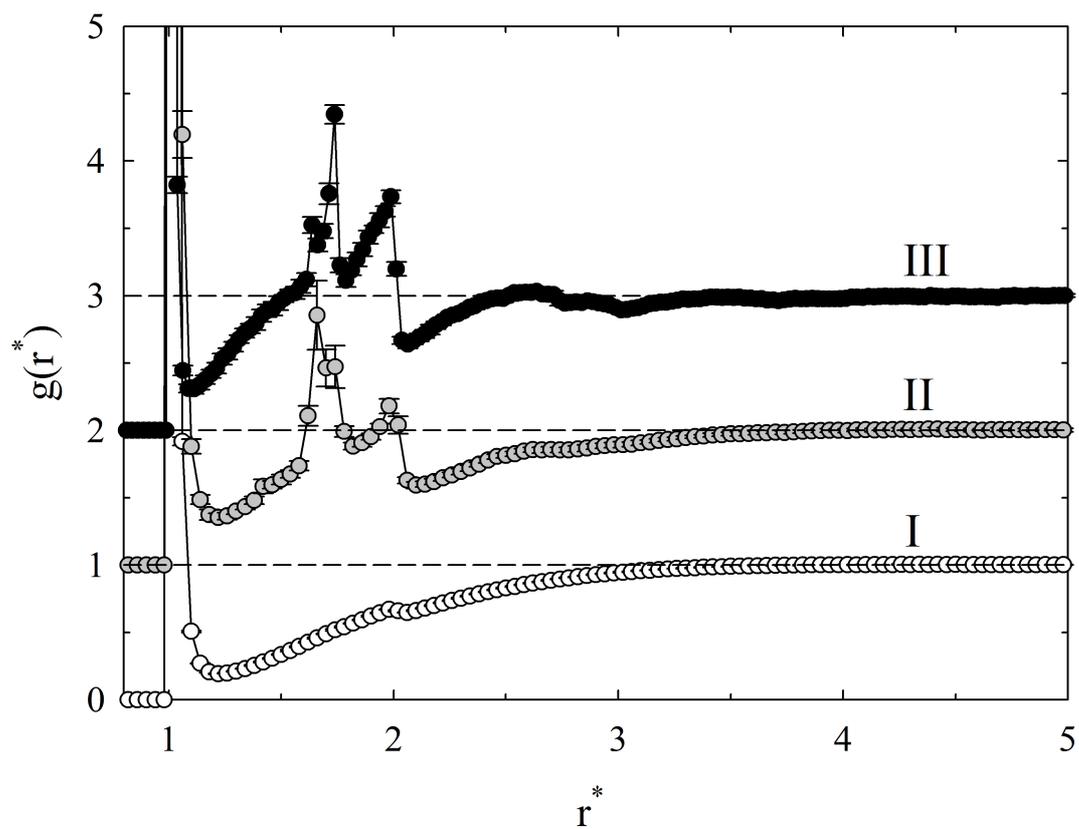

**figure 7.** Radial distribution functions computed for lysozyme at 293 K and pH 7.8. The volume fractions of proteins are: I-0.012, II-0.033, and III-0.12. The distance between proteins is expressed in reduced units, i.e. $r^* = r \cdot \sigma^{-1}$, where $\sigma = 3.648$ nm.